\documentclass[aps,prl,reprint,groupedaddress]{revtex4-1}

\usepackage{amsmath}            
\usepackage{amssymb}            
\usepackage{latexsym}
\usepackage{subfigure}
\usepackage{rotating}
\usepackage{graphicx}
\begin{document}


\title{Granular Dynamics during Impact}


\author{K. N. Nordstrom$^1$, E. Lim$^{2}$, M. Harrington$^{1}$, W. Losert$^1$}
\affiliation{$^1$Institute for Physical Science and Technology, and Department of Physics, University of Maryland, USA}
\affiliation{$^2$Department of Physics and Astronomy, Duke University, USA}


\date{\today}

\begin{abstract}
We study the impact of a projectile onto a bed of 3 mm grains immersed in an index-matched fluid. We vary the amount of prestrain on the sample, strengthening the force chains within the system. We find this affects only the prefactor of the linear depth-dependent term in the stopping force. We propose a simple model to account for the strain dependence of this term, owing to increased pressure in the pile. Interestingly, we find that the presence of the fluid does not affect the impact dynamics, suggesting that dynamic friction is not a factor. Using a laser sheet scanning technique to visualize internal grain motion, we measure the trajectory of each grain throughout an impact.  Microscopically, our results indicate that weaker initial force chains result in more irreversible, plastic rearrangements, suggesting static friction between grains does play a substantial role in the energy dissipation. 
\end{abstract}

\pacs{81.05.Rm 81.40.Np 81.70.Bt}

\maketitle

Take a run on the beach, and your foot strikes a granular material in much the same way an asteroid hits a planet. Even though this interaction is commonplace, the physics of it remain largely mysterious. Previous work has mostly focused on the intruder's dynamics and has focused on continuum models, offering little insight into the microscopics, where the actual grain rearrangments take place  \cite{Katsuragi2007,Allen1957,Tsimring2008,Walsh2004,Hou2005,Goldman2008,Pacheco2011,Ambroso2005_1,Ambroso2005_2}. It's not surprising the microscopics are ignored; granular materials are difficult to study. In addition to being opaque, granular materials are not well-behaved: they are heterogeneous, can behave either like liquids or solids, and exhibit shear localization and jamming phenomena \cite{Jaeger1996}. To make predictions on practical applications of granular impact, such as how far an asteroid will penetrate into soil, we first need a more complete understanding of the entire system. 

When an intruder goes into a granular material, the material behaves somewhat counterintuitively. The material exerts a stopping force that increases as the impact energy (in this case, drop height) increases. One consequence of this behavior is a a typical scaling law found:

\begin{equation}
d\sim H^{1/3}
\label{dvsH}
\end{equation}

\noindent where $d$ is the penetration depth and $H$ is the $total$ drop height, the initial height above the bed plus the penetration depth \cite{Ambroso2005_1}. As the impact energy is increased, it is dissipated over a (relatively) shorter distance. However, this scaling does not appear to be universal - different studies report different scaling exponents for similar systems, underscoring a need to see into the details \cite{Ambroso2005_1}. Further, much work has focused on the same initial state: loose granular matter. More recent work has shown that controlled modifications to the packing fraction, cohesion, and interstitial gas produce different outcomes \cite{Royer2011, Marston2012, Umbanhowar2012}.

Katsuragi and Durian \cite{Katsuragi2007} interpreted the depth scaling as the result of an empirically-determined force law: 

\begin{equation}
\Sigma F = -mg+k|z|+mv^2/d_1
\label{force}
\end{equation}

\noindent where $k$ and $d_1$ are constants obtained by fits. The $mv^2/d_1$ term is due to inertial drag, arising from momentum transfer during grain-intruder and grain-grain collisions. The form of this force law is agreed upon for shallow impacts, though the rate-independent $k|z|$ term saturates with deeper impacts \cite{Pacheco2011}. This term is not due to friction between the grains and intruder \cite{Albert}. Rather, the $kz$ term is generally agreed to be an effective frictional force on the intruder due to the depth-dependent pressure on the intruder \cite{Hou2005,Seguin2009,Katsuragi2007,Brzinski2010,BrzPRL}. The intruder slows down due to forces exerted normally by the surrounding grain network, which sum to give an upward force \cite{BrzPRL}. Interestingly, these forces appear to be transient, with pulses traveling acoustically through force chains \cite{Clark2012}.

A remaining question is how much of a role granular friction plays in these dynamics. Hou \cite{Hou2005} found that depth-dependent term is the correct order of magnitude consistent with a simple hydrodynamic pressure on the intruder, with no need to incorporate the internal friction of a granular material. Seguin \cite{Seguin2009}, in a simulation, reproduced impact scaling with no friction between grains. But Durian et. al. \cite{BrzPRL,Katsuragi2007} find the term to be an order of magnitude larger than the pressure would suggest, and dependent on friction. In another simulation, Tsimiring and Volfson \cite{Tsimring2008}  found that grain-grain friction was responsible for most of the energy dissipation, indicating that in real systems, friction ought to be important.


In this Letter, we vary the initial state by pre-straining the sample, in order to vary the strength of force chains within the sample. Force chains are filamentary networks of contacting grains that bear the weight of the system, transmit energy along their length, and resist shear and buckling due to the friction between grains \cite{Murdoch2013}. We analyze how modifying this network affects the macroscopic scaling of the intruder's dynamics, and thus the stopping force. We then investigate the microscopic failure of the granular material. To do this, we look inside a granular bed during impact, tracking particles near the intruder. 


\begin{figure}[t]
\centering
\includegraphics[width=0.8\columnwidth]{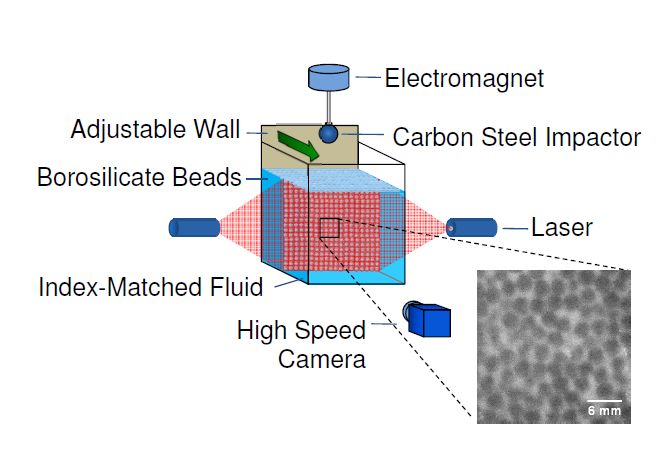}
 \caption{A schematic of the experimental setup. The beads are in an index-matched fluid (DMSO) with a fluorescent dye, allowing illumination of the sample in the impact plane. The sample is strained via displacement of the back wall.}
 \label{fig:Trajectories}
\end{figure}

Our experimental setup is shown in Fig. 1. Our grains are 3 mm glass borosilicate spheres (Glen Mills). They are poured to fill a clear box,  15 cm on each side. The box has one moveable sidewall to apply strain to the system. The box is then filled with a mixture of DMSO, water, and Nile Blue 690 perchlorate dye. The height of the fluid equals the height of the grains, so the intruder does not meet a significant mass of pure fluid. The fluid mixture is tuned to match the index of refraction of the grains; the index mismatch is less than 0.005. A laser sheet illuminates the plane of impact, resulting in bright fluid and dark grains \cite{Dijksman2012}. The laser sheet and moveable sidewall are positioned by stepper motors.

For each impact, we take 2D video data in the impact plane with a high speed (200-1000 fps) PCO.edge camera.  The intruder is a 1 inch carbon steel sphere, released via an electromagnet. The size ratio of container to intruder is sufficient to minimize wall effects \cite{Nelson2008, Stone2004}.

With the moveable sidewall, we build up force chains in the sample by (very slightly) compressing the material from the side. The force chains will then be biased perpendicular to the direction of impact.  During the straining procedure, we keep the surface of the sample level by resting a light, flat plate on the surface of the sample. We start in a packed state, $\phi=0.641\pm0.006$. Due to the presence of the top plate, we may expect slight compaction when straining. This maximum volume fraction increase is approximately 0.006 at a strain of 1\%. This is less than the amounts in \cite{Umbanhowar2012} needed to have an effect on dynamics. So by straining the sample, we create a stronger force chain network only, not a denser packing. After the sample is strained, the top plate is removed, the projectile is dropped into the bed and the impact is recorded. We restore the unstrained state by reversing the compression, and stirring for several minutes.

From the videos we can extract 1) the initial and final position of the intruder, 2) the instantaneous position (and velocity) of the intruder, and 3) the positions (and velocities) of the grains, using our established particle tracking routines \cite{Slotterback2008}.

\begin{figure}[t]
\centering
\includegraphics[width=0.8\columnwidth]{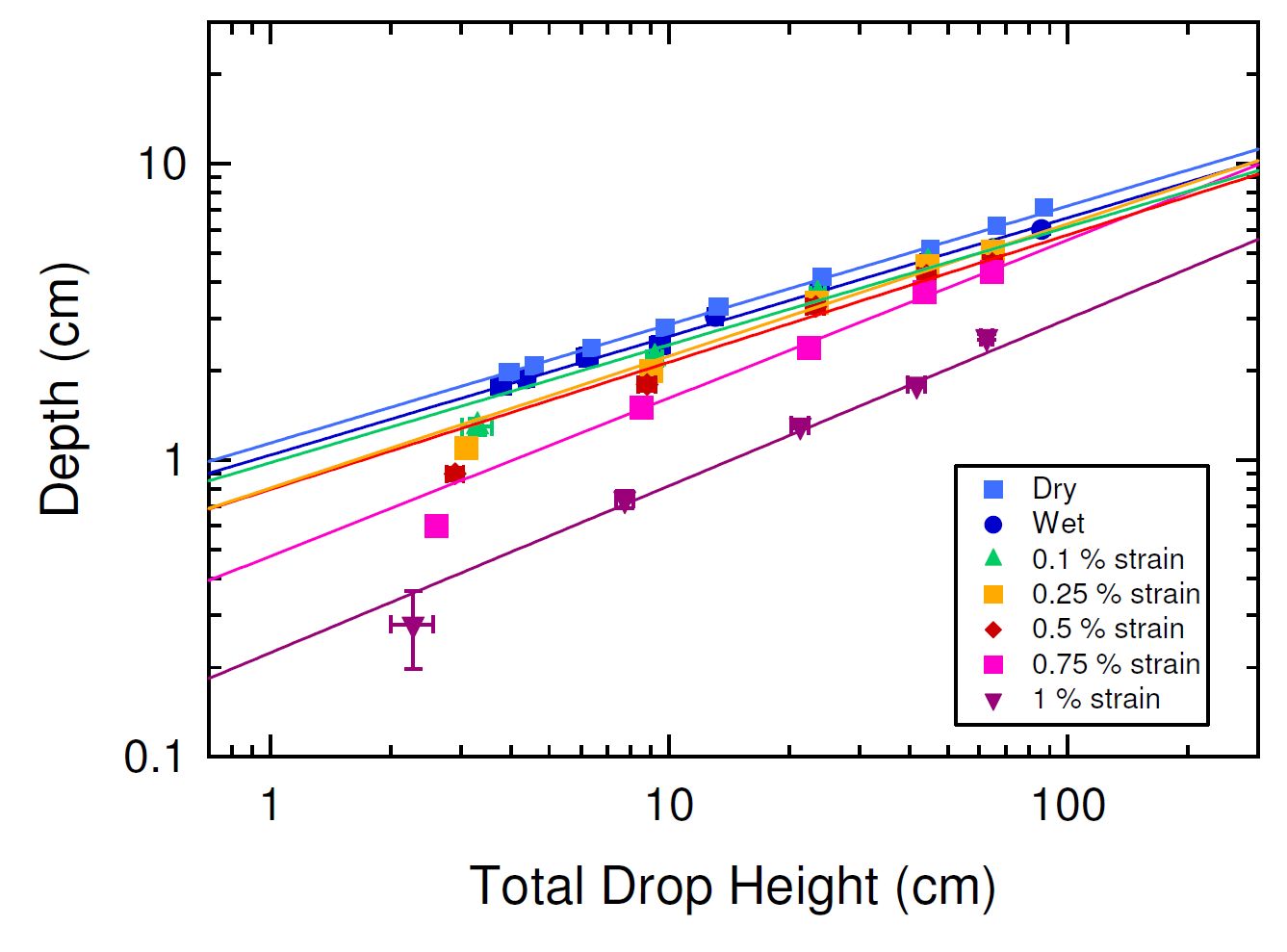}
 \caption{Depth vs total height for many different sample preparations. The fits shown are to simple power laws. Each point corresponds to at least three experimental runs. The dry and wet unstrained samples show the same exponent $\approx$0.4, and differ in prefactor by about 15\%. }
 \label{fig:Trajectories}
\end{figure}

\begin{figure}[t]
\centering
\includegraphics[width=0.75\columnwidth]{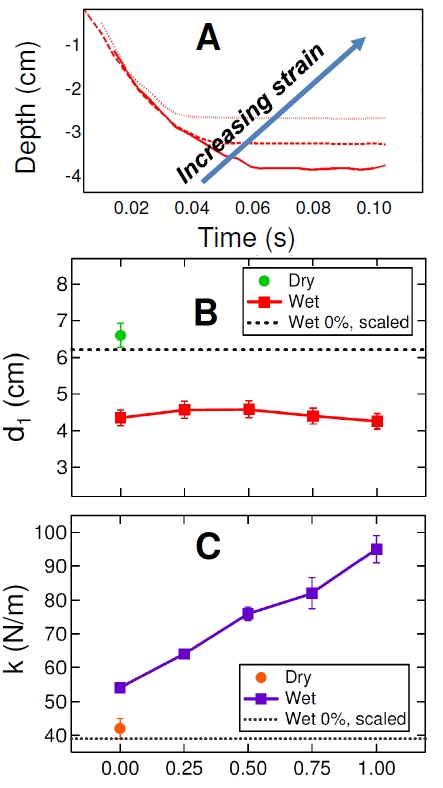}
 \caption{(a) Three sample trajectories 
for a fixed drop height with differing amounts of prestrain. From the 
trajectories, we are able to find a value of (b) $d_1$ for each strain value 
and (c) $k$ for each strain. The dashed and dotted lines show the 
values of $d_1$ and $k$ for the unstrained wet impact, scaling out fluid mass.}
 \label{fig:Trajectories}
\end{figure}

We examine the scaling of $d$, the penetration depth, with the total drop height $H$, in Fig. 2. Each data point corresponds to 5 experiments. For our unstrained system (blue circles), we see that indeed a power law captures the data.  The best-fit exponent is $\approx$0.4, a departure from the simple 1/3 scaling but well within the range of exponents seen for various shallow impacts ($\approx$0.25-0.50). 

We have also looked at our system and compared it to a dry system (blue squares). The depth scaling does not change for wet vs dry. The only difference is the 15\% difference in prefactor.  This is not surprising for two reasons. First, we are in a fully saturated state, so surface tension effects are negligible  \cite{Marston2012}. Second, the Reynolds numbers for the wet and dry systems are both high, well out of any laminar flow regime for fluid in the voids. We have performed auxiliary measurements showing that a viscosity increase of over 100-fold is necessary to appreciably modify the dynamics. Lastly, we can show that the added mass of the fluid is the sole reason for the difference in scaling, as will be detailed later. 

We find that prestrain greatly changes the power law scaling of depth vs. drop height (Fig. 2), with some deviation from a power law at extremely low drop heights. Even small strain results in a much shallower impact. Further, the exponent is increased, though only to a maximum value of $\approx$0.5.  This suggests that with stronger force chains, the impact resistance of the material is less, perhaps because the material is already somewhat hardened.  The direction of deviation of the shallow heights is not surprising; for low energy drops, the surface breaking starts to become much more onerous to the intruder.

Example intruder trajectories are given in Fig. 3a for three different strains. From these trajectories we analyze the stopping force, using the position data and its derivative, avoiding noise from further derivatives. The details of this are in \cite{Clark2013}. We find that Eq. 2 is adequate to describe the motion at all strains. 




We calculate the prefactors $k$ and $d_1$ and see how they depend on strain, as shown in Figs. 3b and 3c. Each point is calculated using data from many drop heights. We see the inertial constant, $d_1$, does not vary with increasing strain. However the friction force constant $k$ monotonically increases with strain. Our value of $d_1$ is factor of 2 smaller than in \cite{Katsuragi2007}, perhaps due to the difference in bead/intruder size ratio. Our (unstrained) value of $k$ is in agreement with \cite{Katsuragi2007}.

Included in Fig. 3b and c is data for the dry unstrained sample. $k$ will scale with the mass surrounding the intruder \cite{Katsuragi2007, Hou2005, Seguin2009} due to its hydrostatic pressure-like form and $d_1$ should scale inversely; more material must be moved \cite{Katsuragi2007}. What follows is: $d_1\sim \frac{d\rho}{\rho_g}$ and $k\sim\frac{mg\rho_g}{d\rho}$, where $\rho_g$ is the density of the granular material, $\rho$ is the density of the intruder, and $d$ is the intruder diameter \cite{Seguin2009}. When rescaled with the added weight of the wet system ($\rho_{wet}=1.82$ g/cc, $\rho_{dry}=1.43$ g/cc), we see $k$ and $d_1$ are independent of fluid immersion (Figs. 3b,c).


However, the interstitial fluid might also provide lubrication, thus changing the dynamics. We have performed two measurements of friction with our beads. We find similar static coefficients of friction $\mu_{dry}=0.36\pm0.02$ and $\mu_{wet}=0.39\pm0.02$ by testing the angle of repose. But a test of the dynamic friction tells a different story: $\mu_{dry}=0.35\pm0.02$ and $\mu_{wet}=0.26\pm0.02$. If the dynamic friction influences the $kz$ term at all, then we would expect $k$ (adjusted for the mass of the fluid) to be smaller for the wet case, and we find them to be the same (see Fig. 3c, dotted line). Tsimring and Volfson \cite{Tsimring2008} found grain-grain friction was important, but did not distinguish between static and dynamic friction. Our results present a strong case that if $k$ is influenced by friction, it is primarily static friction that matters. As static contacts in force chains are enhanced by the prestrain, this picture aligns with the increase in $k$. 

We have shown that prestrain, which enhances the force chains between particles, leads to an increased $kz$ term in the stopping force. Brzinski et. al. \cite{BrzPRL} suggest:
\begin{equation}
k=\alpha \mu \rho_g  g A.
\label{kscaling}
\end{equation}

\noindent where $A$ is the cross-sectional area of the intruder. $\mu \rho_g  g A$ is what might be expected at face value, but the prefactor $\alpha$ is of order 20. Eq. 3 is consistent with our results, and we find $\alpha$ is 16 for the zero strain case and 27 for the highest. That $\alpha>1$ is explained in the context of force chains extending into the sample. More contacts than those at the intruder's surface are participating in slowing it down. $\alpha$ then possibly represents an effective size of the network. Conversely, both Seguin \cite{Seguin2009} and Hou \cite{Hou2005} find that $k\approx\rho_g  g A$, identical scaling to Eq. 3, but with no need for the friction coefficient or a prefactor. 

Seguin's simulation included no interparticle friction, but even frictionless beads have internal friction angles \cite{static}. So if we take $\alpha=20$ and $\mu=0.05$ \cite{static}, the expression from \cite{BrzPRL} appears to be more general, and not in conflict. (It should be noted that Seguin concedes that the quantitative details might depend on friction.) The value of $k$ found by Hou \cite{Hou2005} might still not be predicted by Eq. 3, but their $\mu$ is unknown. Further, their beads were very light ($\rho_g = 0.37$ g/cc), it's plausible that the contact network was very short-ranged, resulting in $\alpha\mu\approx1$.

Thus, Eq. 3 seems satisfactory for the unstrained case, but is not general enough to describe our experiments. We propose the an additional factor that incorporates the increased strength of the material. It is a quantity that is equal to 1 when grain-grain contacts are in their ``natural'' state, and increases as the local pressure increases. The excess pressure is an established function of uniaxial strain \cite{Nordstrom}: $P= \frac{\phi Z E}{6 \pi (1-\nu^2)}\gamma^{3/2}=Y\gamma^{3/2},$ where $Z$ is the average contact number, $E$ is the grain elastic modulus, and $\nu$ is their Poisson ratio, and $Y$ is a constant with units of pressure encapsulating these parameters. Thus, we propose a modification to Eq. 3:

\begin{equation}
k= \alpha \mu \rho_g  g A(1+\frac{Y}{Y_0}\gamma^{3/2}),
\label{modified}
\end{equation}

\noindent where $Y_0$ is then a reference pressure characteristic of the system. Thus $Y/Y_0$ represents the strength of contact enhancement. This expression reduces to Eq. 3 when the strain is zero and monotonically increases with strain. Using our data we find $Y_0$ to be about 15 MPa, close to what we would expect for the elastic modulus of a glass sphere pack \cite{glass}. (We do not attempt to calculate this from first principles as this is an unresolved question \cite{vanhecke}.)

\begin{figure}[b]
\centering
\includegraphics[width=0.8\columnwidth]{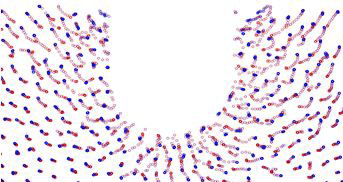}
 \caption{Particle tracks of the grains during impact. The tracks go from red (early) to blue (late). Grains below the intruder move down, and grains to the side of the intruder move out and up. The frame-to-frame motion is not always smooth, due to the frustrated nature of the packing. }
 \label{fig:Trajectories}
\end{figure}

Finally, we look into the grain-grain energy dissipation. There are 3 modes to consider: restitutional losses, frictional losses, and force chain splitting \cite{Clark2012}. We cannot easily look into the restitutional losses or characterize each chain, but can look for frictional signatures at the particle scale: force chain buckling.


\begin{figure}[t]
\centering
\includegraphics[width=1\columnwidth]{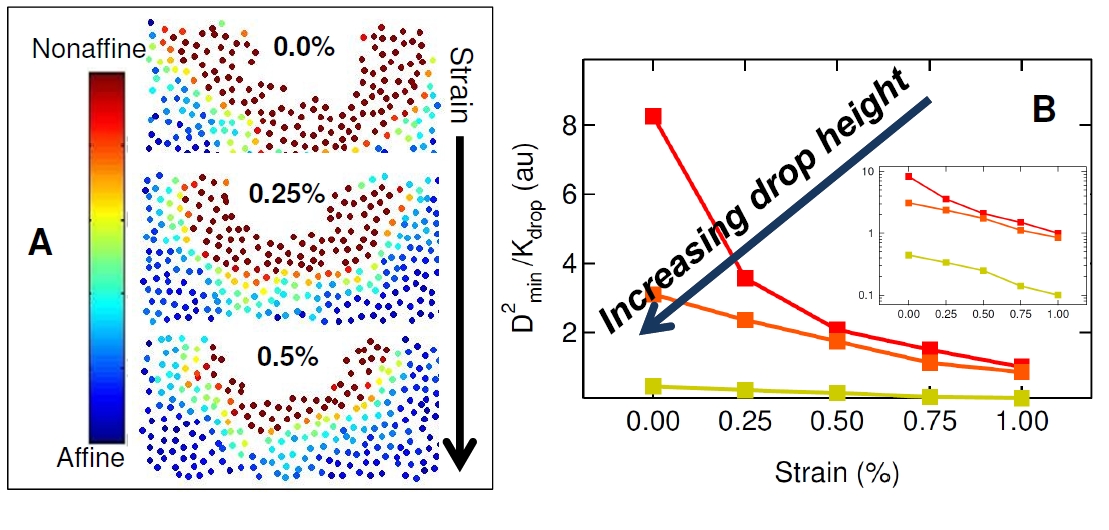}
 \caption{(a) Spatial map of $D^2_{min}$ for three different strains. The drop height is the same in each case, and $D^2_{min}$ is measured in the $z=0$ limit. (b) For 3 drop heights and 5 strains, the average $D^2_{min}$ value is normalized by the energy of the intruder just before impact.}
 \label{fig:Trajectories}
\end{figure}

By looking at the trajectories of the particles, we can gain further insight into energy dissipation at the microscale. An example tracked data set is shown in Fig. 4. We see the trajectories are characterized by downward motion below the impactor, and upwards and outwards motion to the side. This flow field has been characterized before only in 2D systems, but we find it in good qualitative agreement with them \cite{Clark2012,Kondic2012}.

We then measure the signatures of buckling force chains \cite{Murdoch2013,Makse1999,Tordesillas2008}: local plastic rearrangements of particles, using the quantity $D^2_{min}$ [26-28]:

\begin{equation}
D^2_{min,i}=\text{min}\lbrace\sum_j[\Delta\overline d_{ij}(t)-E_i\overline d_{ij}]\rbrace ^2
\label{D2min}
\end{equation}

$D^2_{min,i}$ quantifies the nonaffine motion of $j$ particles in the neighborhood around a given particle $i$ after removing the averaged linear response to the strain, given by tensor $E_i$; a larger $D^2_{min}$ indicates more nonaffine motion. The vector $\overline d_{ij}$ is the relative position of $i$ and $j$,  $\Delta\overline d_{ij}$ is the relative displacement after a delay time $\Delta t$, 5 ms in this case. The neighborhood radius around a particle is 2.2 particle diameters, capturing  $\approx$10-15 particles.

For the initial impact ($z=0)$, we find a high $D^2_{min,i}$ zone near the projectile (Fig. 5a); this result is in agreement with 2D simulations of impact in \cite{Kondic2012}. By looking at the average value of $D^2_{min,i}$ in our field of view, we can compare how the plastic rearrangements (nonaffine motion) varies with initial sample strain (Fig. 5b). Increasing the initial strain results in decreased nonaffine motion. This suggests that the buckling of force chains decreases as the chains get stronger, suggesting friction is responsible for some microscale dissipation, as in \cite{Tsimring2008}. Indeed Kondic et. al. \cite{Kondic2012} found that by introducing friction, the nonaffine motion was decreased. 


In Fig. 5b we also show that $D^2_{min}$, when properly scaled by the impact energy, is most disturbed by strain for low-energy impacts. This lends further support to the notion that microscale friction is important - at lower impact energies static friction should be more important. And by increasing the static friction, the plastic response signifying network breakage decreases.  

As an entire picture, it seems the intruder slows down due to normal forces exerted (transiently) by the force chains in the pile. Strengthening the chains enhances this deceleration, underscoring the importance of this network. This is the first study of uniaxial strain on such a  system. We also propose an expression for $k$ that incorporates the fabric of the material. This can be further tested by studying a variety of materials.

By comparing wet and dry systems, we see it is the static frictional contact network that determines the dynamics, not the dynamic friction between grains. By measuring bead-scale motions, we see that friction plays a role in the grain-grain dissipation via force chain buckling; the nonaffine motion trends support this. What is likely also occurring are restitutional losses and force chain splitting. Future work should modify the friction between grains or their softness to isolate these effects. 
\begin{acknowledgements}
We acknowledge support from U.S. DTRA under
Grant No. HDTRA1-10-0021. We thank Don Martin and Steve Slotterback for technical support, and Bob Behringer, Abe Clark, and Lou Kondic for helpful discussions. 
\end{acknowledgements}
\bibliography{impactrefs}

\end{document}